\documentclass[11pt]{article}
\usepackage[a4paper,hmargin=1.0in,vmargin=1.0in]{geometry}
\usepackage{mathtools,amsthm,amssymb,setspace,sectsty,bbm}
\usepackage[round]{natbib}
\usepackage[usenames,dvipsnames]{xcolor}
\usepackage[linktocpage=true,pagebackref=true,colorlinks,allcolors=blue,bookmarks=true,bookmarksopen,bookmarksnumbered]{hyperref}

\newtheoremstyle{DStheorem}
  {\topsep}
  {\topsep}
  {\itshape}
  {0pt}
  {\scshape}
  {.}
  { }
  {\thmname{#1}\thmnumber{ #2}\thmnote{ (#3)}}
\theoremstyle{DStheorem}
\newtheorem{theorem}{Theorem}[section]
\newtheorem{lemma}[theorem]{Lemma}
\newtheorem{claim}[theorem]{Claim}
\newtheorem{observation}[theorem]{Observation}
\newtheorem{corollary}[theorem]{Corollary}

\let\oldproofname=\proofname
\renewcommand{\proofname}{\rm\sc{\oldproofname}}

\newcommand{\MyAbove}[2]{\genfrac{}{}{0pt}{}{#1}{#2}}
\newcommand{\bs}[1]{\boldsymbol{#1}}
\newcommand{\bbR}{\mathbbm{R}}
\newcommand{\eps}{\epsilon}

\newcommand{\opt}{\mathrm{OPT}}
\newcommand{\cc}{\mathrm{CC}}

\sectionfont{\large} \subsectionfont{\normalsize}
\allowdisplaybreaks
\makeindex

\begin{document}

\begin{titlepage}

\title{The Continuous-Time Joint Replenishment Problem: \\
$\eps$-Optimal Policies via Pairwise Alignment}
\author{%
Danny Segev\thanks{Department of Statistics and Operations Research, School of Mathematical Sciences, Tel Aviv University, Tel Aviv 69978, Israel. Email: {\tt segevdanny@tauex.tau.ac.il}. Supported by Israel Science Foundation grant 1407/20.}}
\date{}
\maketitle

\setcounter{page}{200}
\thispagestyle{empty}

\begin{abstract}
The main contribution of this paper resides in developing a new algorithmic approach for addressing the continuous-time joint replenishment problem, termed $\Psi$-pairwise alignment. The latter mechanism, through which we synchronize multiple Economic Order Quantity models, allows us to devise a purely-combinatorial algorithm for efficiently approximating optimal policies within any degree of accuracy. As a result, our work constitutes the first quantitative improvement over power-of-$2$ policies, which have been state-of-the-art in this context since the mid-80's. Moreover, in light of recent intractability results, by proposing an efficient polynomial-time approximation scheme (EPTAS) for the joint replenishment problem, we resolve the long-standing open question regarding the computational complexity of this classical setting.
\end{abstract}


\end{titlepage}

\setcounter{page}{200}
\thispagestyle{empty}
\tableofcontents

\newpage
\setcounter{page}{1}
\section{Introduction}

Dating back to the mid-60's, with indirect investigations surfacing even earlier, the joint replenishment problem has been playing an instrumental role in developing the theoretical foundations of inventory management as well as in boosting the  practical appeal of this academic field. Given the massive body of work along these lines, spanning rigorous methods, heuristics, experimental studies, real-life applications, and software solutions, we cannot do justice and present an exhaustive overview of this literature beyond directly-related results. We refer avid readers to excellent survey articles \citep{AksoyE88, GoyalS89, MuckstadtR93, KhoujaG08, BastosMNMC17} and book chapters \citep{SilverP85, Zipkin00, MuckstadtS10}, as well as to the references therein, for an in-depth discussion in this context.

By consulting the above-mentioned resources, it is apparent that while joint replenishment settings have been studied in various forms and shapes, they are all inherently concerned with the lot-sizing of multiple commodities over a given planning horizon. Here, our seemingly-simple objective is to determine a joint replenishment policy that minimizes long-run average operating costs. However, these circumstances give rise to unique computational obstacles regarding the efficient synchronization of numerous Economic Order Quantity (EOQ) models along with wide-open analytical questions regarding their structural characterization. In a nutshell, on top of commodity-specific ordering and inventory holding costs, what makes such synchronization particularly challenging is the interplay between different commodities via joint ordering costs, incurred whenever an order is placed, regardless of its contents. To take a deeper dive into these questions and to accurately position the main contributions of our work, we proceed by providing a formal mathematical description of the joint replenishment problem in its broadest continuous-time form.

\subsection{Model formulation} \label{subsec:model_definition}

\paragraph{The Economic Order Quantity (EOQ) model.} For ease of exposition, it is convenient to start off by describing the basic building block of joint replenishment settings: The Economic Order Quantity model. Here, we wish to determine the time interval $T$ between successive orders of a single commodity, aiming to minimize our long-run average cost over the continuous planning horizon $[0,\infty)$. Specifically, this commodity is associated with a stationary demand rate of $d$, to be fully satisfied upon occurrence, meaning that lost sales and back orders are not permitted. In this context, periodic policies are simply those where the ordering frequency is uniform, namely, orders will be placed at time points $0, T, 2T, 3T, \ldots$, where the time interval $T$ is a decision variable to be optimized. In turn, each of these orders incurs a fixed cost of $K$, regardless of its quantity. This ingredient is complemented by a linear holding cost of $h$, incurred per time unit for each inventory unit in stock. 

Our objective is to determine the time interval $T$ so as to minimize long-run average ordering and holding costs. Based on this description, one can easily verify that, in an optimal policy, orders will be placed only when the current on-hand inventory drops to zero, i.e., zero inventory ordering (ZIO) policies are optimal. As such, the objective function of interest admits a succinct representation, given by
\[ C(T) ~~=~~ \frac{ K }{ T } + HT \ , \]
with the convention that $H = \frac{ hdT }{ 2 }$. The next claim summarizes a number of very well-known properties exhibited by this function; its derivation requires elementary arguments and can be found in any relevant textbook.

\begin{claim} \label{clm:EOQ_properties}
The cost function $C : (0,\infty) \to \bbR_{++}$ satisfies the following properties:
\begin{enumerate}
    \item $C$ is strictly convex.

    \item The unique minimizer of $C$ is $T^* = \sqrt{ K / H }$.

    \item $C( \theta \cdot T^* ) = \frac{ 1 }{ 2 } \cdot (\theta + \frac{ 1 }{ \theta } ) \cdot C( T^* )$, for every $\theta > 0$.
\end{enumerate}
\end{claim}

\paragraph{The joint replenishment problem.} With these preliminaries in place, an instructive way of viewing the joint replenishment problem is through the following question: How should we synchronize numerous Economic Order Quantity models, when different commodities are interacting via joint ordering costs? Specifically, we have at our possession a set of $n$ commodities, where each commodity $i \in [n]$ is associated within its own EOQ model, parameterized by ordering and holding costs $K_i$ and $H_i$, respectively. Here, a time interval $T_i$ between successive
orders of this commodity leads to marginal operating costs of the form
\[ C_i(T_i) ~~=~~ \frac{ K_i }{ T_i } + H_i T_i \ . \]
However, as previously mentioned, further complexity comes in the form of a joint ordering cost, $K_0$, incurred whenever an order is placed, regardless
of its particular subset of commodities.

Given these ingredients, a joint replenishment policy can be represented as a vector $T = (T_1, \ldots, T_n)$, specifying the ordering interval $T_i$ of each commodity $i \in [n]$. For such policies, the first part of our objective function corresponds to the sum of marginal EOQ-based costs, $\sum_{i \in [n]} C_i( T _i )$. The second part, which will be designated by $J(T)$, captures long-run average joint ordering costs. This term can be formally expressed as
\begin{equation} \label{eqn:joint_cost_def}
J( T ) ~~=~~ K_0 \cdot \lim_{\Delta \to \infty} \frac{ N(T,\Delta) }{ \Delta } \ ,
\end{equation}
where $N(T,\Delta)$ stands for the number of joint orders in $[0,\Delta]$ with respect to the time intervals $T_1, \ldots, T_n$. That is, letting ${\cal M}_{T_i,\Delta} = \{ 0, T_i, 2T_i, \ldots, \lfloor \frac{ \Delta }{ T_i } \rfloor \cdot T_i \}$ be the set of integer multiples of $T_i$ within $[0,\Delta]$, we  have
\[ N(T,\Delta) ~~=~~ \left| \bigcup_{i \in [n]} {\cal M}_{T_i,\Delta} \right| \ . \]
Consequently, our objective is to identify a joint replenishment policy $T = (T_1, \ldots, T_n)$ that minimizes long-run average operating costs, represented by
\[ F(T) ~~=~~ J(T) + \sum_{i \in [n]} C_i( T_i ) \ . \]

\paragraph{The joint ordering term $\bs{J(T)}$.} We mention in passing that, by glancing at equation~\eqref{eqn:joint_cost_def},
it may not be clear why $\lim_{\Delta \to \infty} \frac{ N(T,\Delta) }{ \Delta }$ necessarily exists for any given policy $T$. To address this question, for any subset ${\cal N} \subseteq [n]$, let $M_{\cal N}$ be the least common multiple of $\{ T_i \}_{i \in {\cal N}}$, with the convention that $M_{\cal N} = \infty$ when these time intervals do not have common multiples. In Section~\ref{subsec:proof_lem_limit_joint_order}, we prove the next claim, showing that the above-mentioned limit indeed exists, along with an explicit expression. That said, the latter consists of exponentially-many terms, implying that merely computing the joint ordering cost of an arbitrarily-structured policy may not be trivial.

\begin{lemma} \label{lem:limit_joint_order}
$\lim_{\Delta \to \infty} \frac{ N(T,\Delta) }{ \Delta } = \sum_{{\cal N} \subseteq [n]} \frac{ (-1)^{ |{\cal N}| + 1 } }{ M_{\cal N} }$.
\end{lemma}

\subsection{Known results and open questions} \label{subsec:related_work}

In what follows, we highlight cornerstone results regarding rigorous algorithmic methods for efficiently identifying provably-good joint replenishment policies. As previously mentioned, further background on historical developments, including classical paper such as those of \cite{Zangwill66}, \cite{Veinott69}, and \cite{Kao79}, can be attained by consulting directly-related surveys and books.

\paragraph{Power-of-$\bs{2}$ policies.} To our knowledge, the seminal works of \citet{Roundy85, Roundy86}, \cite{JacksonMM85}, \cite{MaxwellM85}, and \cite{MuckstadtR87} were the first to obtain efficiently-constructible and provably-good performance guarantees for the joint replenishment problem. Since then, these findings have become some of the most renowned breakthroughs in inventory management, perhaps due to their applicability in a wide range of lot-sizing problems. An in-depth discussion of these developments is provided, for example, by \cite{MuckstadtR93}. Moreover, for very elegant methods of deriving these results, we refer readers to the work of \cite{TeoB01} in this context.

In a nutshell, the above-mentioned papers propose various methods for rounding optimal solutions to convex relaxations, ending up with so-called power-of-$2$ policies. The latter fix a common base, $T_0$, with each time interval $T_i$ being of the form $2^{ \mu_i } \cdot T_0$, for some integer $\mu_i$. Surprisingly, when $T_0$ is specified in advance, this ingenious mechanism for synchronizing joint orders determines a corresponding power-of-$2$ policy whose long-run average cost is within factor $\sqrt{9/8} \approx 1.06$ of optimal. Even more surprisingly, when $T_0$ can be optimized, the latter factor can be sharpened to $\frac{ 1 }{ \sqrt{2} \ln 2 } \approx 1.02$, constituting the best known approximation guarantee for the joint replenishment problem to this day. Consequently, as stated in countless papers, books, conference talks, and course materials, the primary open questions that motivate our work can be briefly summarized as follows:
\begin{quote}
{\em Can we devise stronger mechanisms for synchronizing multiple EOQ models, possibly outperforming power-of-$2$ policies? Could these ideas be leveraged to attain improved approximation guarantees for the joint replenishment problem?}
\end{quote}

\paragraph{Intractability results.} Digging from the other end of the tunnel, exciting progress has been made in the last decade regarding the plausibility of efficiently computing optimal policies. This line of work was initiated by \cite{SchulzT11}, establishing intricate connections between this question and fundamental problems in number theory. Specifically, they proved that when all time intervals are required to be multiples of a prespecified common base, a polynomial time algorithm for the joint replenishment problem would imply an analogous result for integer factorization. Subsequently, by exploiting the extraordinary work of \cite{Zhang14} on bounded gaps between successive primes, \cite{CohenHillelY18} proved that the fixed-base version is in fact strongly NP-hard. The latter result has been considerably streamlined by \cite{SchulzT22}, showing that NP-hardness arises even in the presence of only two commodities. Finally, for the joint replenishment problem with arbitrarily-structured periodic policies, which is precisely the topic of our work, \cite{SchulzT22} have recently extended their original findings to derive its polynomial-relatability to integer factorization. The latter result was further lifted to a full-blown strong NP-hardness proof by \cite{TuisovY20}. Given this state of affairs, yet another repeatedly-occurring open question is whether the above-mentioned evidence for intractability is best-possible or not, or put differently:
\begin{quote}
{\em Is the joint replenishment problem APX-hard? Alternatively, could this setting admit an approximation scheme?}    
\end{quote}

\paragraph{Earlier attempts.} Interestingly, over the years, a handful of authors have been successful at deriving approximation schemes for several different variants of the joint replenishment problem. These efforts seem to have been initiated by \cite{LuP94}, who designed a fully polynomial-time approximation scheme (FPTAS) within the class of integer-ratio policies. Nevertheless, as demonstrated by \citet{Roundy85, Roundy86}, there is still a multiplicative gap of $\frac{ 1 }{ \sqrt{2}  \ln 2 } \approx 1.02$ between the best such policy and an arbitrarily-structured one, meaning that this approach has an inherent constant-factor loss in optimality. More recently, approximation schemes have also been derived for the joint replenishment problem in its discrete-time finite-horizon setting, initially for periodic policies \citep{SchulzT11}, and subsequently for non-periodic ones \citep{Segev14, NonnerS13}. That said, limiting attention to an evenly-spaced finite horizon goes around dealing with most synchronization issues arising due to the joint ordering term $J(T)$, whose rather convoluted structure was discussed in Section~\ref{subsec:model_definition}. In essence, such settings leave us with the much simpler task of coordinating a finite number of prespecified integer-valued time periods, and unfortunately, we are unaware of any way to leverage these ideas, so that one attains any non-trivial approximation for the classical joint replenishment problem.

\subsection{Main results}

The primary contribution of this paper resides in developing a new algorithmic approach for addressing the continuous-time joint replenishment problem, termed $\Psi$-pairwise alignment, through which we devise a purely-combinatorial algorithm for efficiently approximating optimal policies within any degree of accuracy. As formally stated in Theorem~\ref{thm:main} below, $\Psi$-pairwise alignment allows us to determine a replenishment policy whose long-run average cost is within factor $1 + \eps$ of optimal. Quite surprisingly, as far as running time is concerned, these ideas fall within the notion of an
efficient polynomial-time approximation scheme (EPTAS), where $\eps$-related terms are separated from input-related terms. 

\begin{theorem} \label{thm:main}
For any $\eps \in (0, \frac{ 1 }{ 2 })$, the classical joint replenishment problem can be approximated within factor $1 + \eps$ of optimal. The running time of our algorithm is $O( 2^{ \tilde{O}(1/\eps^3) } \cdot n^{ O(1) } )$.
\end{theorem}

\paragraph{Comparison to power-of-$\bs{2}$ policies.} Circling back to the list of open questions in Section~\ref{subsec:related_work}, our work constitutes the first quantitative improvement over power-of-$2$ policies, which have been state-of-the-art in terms of provable performance guarantees since the mid-80's. In fact, we believe that the technical ideas behind $\Psi$-pairwise alignment and their analysis could provide approximation schemes for a host of additional inventory management settings; some of these potential applications are discussed in Section~\ref{sec:conclusions}. From a purely practical standpoint, power-of-$2$ policies still appear to be indispensable at present time in terms of simplicity and implementability. However, it is worth mentioning that we have not made any attempt to optimize the running time expression stated in Theorem~\ref{thm:main}. In fact, we preferred to arrive at the simplest and easy-to-understand presentation, leaving quite a bit of room to spare by straightforwardly implementing several algorithmic steps. That said, Section~\ref{sec:conclusions} will briefly touch on possible avenues for more efficient implementations.

\paragraph{Computational characterization of the joint replenishment problem.} Complementing previously-mentioned  intractability results due to \cite{SchulzT22} and \cite{TuisovY20}, our approximation scheme for the joint replenishment problem  resolves the long-standing open question regarding the computational complexity of this setting. As previously mentioned, the work of \cite{TuisovY20} classifies  joint replenishment as being strongly NP-hard, thereby ruling out the potential existence of a fully polynomial-time approximation scheme (FPTAS). Consequently,  due to the inevitable exponential dependency on the accuracy level $\eps$, Theorem~\ref{thm:main} provides the best possible form of such dependency, up to perhaps replacing $2^{ \tilde{O}(1/\eps^3) }$ by lower-order exponential terms.

\paragraph{Outline.} Moving forward, Section~\ref{sec:algorithm} starts off the technical part of this paper by providing a detailed account of our algorithmic approach. Subsequently, Section~\ref{sec:analysis} is dedicated to analyzing our resulting policy, showing that its long-run average cost is within factor $1 + \eps$ of optimal. Along this presentation, several technical arguments and auxiliary claims will be deferred to Section~\ref{sec:more_proofs}, mainly for ease of exposition.
\section{Algorithmic Overview: \texorpdfstring{$\bs{\Psi}$}{}-Pairwise Alignment} \label{sec:algorithm}

In what follows, we lay down the specifics of our algorithmic approach, leaving its performance analysis to be separately discussed in subsequent sections. For this purpose, in Section~\ref{subsec:two_regimes}, we begin by identifying a parametric regime where the joint replenishment problem can easily be handled. This preliminary step enables us to bypass certain irregularities while treating the complementary regime, which captures most technical hurdles. Next, Sections~\ref{subsec:alignment_intuition} and~\ref{subsec:alg_definitions} introduce the overall intuition behind $\Psi$-pairwise alignment and present the basic objects that will be manipulated along with their related terminology. With these ingredients in place, Sections~\ref{subsec:alg_alignment} and~\ref{subsec:structure_feasible} are intended to describe how such alignment can be efficiently computed as well as to highlight some of its structural properties, turning out to be very useful for analytical purposes. Finally, Section~\ref{subsec:policy_description} explains how joint orders and commodity-specific ones will be placed with respect to a given $\Psi$-pairwise alignment, collectively forming our approximate replenishment policy.

\subsection{Two parametric regimes} \label{subsec:two_regimes}

Let us make use of $T^* = (T_1^*, \ldots, T_n^*)$ to denote an optimal replenishment policy, that will be fixed from this point on, with $T_{\min}^* = \min_{i \in [n]} T_i^*$ being the minimal ordering interval of any commodity. We begin by computing an over-estimate $\widetilde{\opt}$ for the optimal long-run average cost $F(T^*)$, such that $F(T^*) \leq \widetilde{\opt} \leq 2 \cdot F(T^*)$. This estimate can be obtained in polynomial time via power-of-$2$ policies, for example, as explained in Section~\ref{subsec:related_work}. For convenience, we plug-in an approximation guarantee of $2$ rather than $\frac{ 1 }{ \sqrt{2} \ln2 } \approx 1.02$, noting that the specific constant does not play an important rule.

\paragraph{The easy regime: $\bs{T_{\min}^* \geq \frac{ n }{ \eps } \cdot \frac{ K_0 }{ \widetilde{\opt} }}$.} In what follows, we argue that when the minimal ordering interval $T_{\min}^*$ is much larger than $\frac{ K_0 }{ \widetilde{\opt} }$, a very simple replenishment policy is actually near-optimal. To this end, let $\hat{T} = (\hat{T}_1, \ldots, \hat{T}_n)$ be the policy where, for each commodity $i \in [n]$, we pick its interval $\hat{T}_i$ as an optimal solution to a standard single-commodity EOQ model (see Section~\ref{subsec:model_definition}), in which the ordering and holding cost parameters are given by $K_0 + K_i$ and $H_i$, respectively. Namely, we overload the joint ordering cost $K_0$ into each commodity-specific order, meaning that the long-run average cost function in this case becomes $C_i^+( T_i ) = \frac{ K_0 + K_i }{ T_i } + H_i T_i$. Accordingly, we have $\hat{T}_i = \sqrt{ (K_0 + K_i) / H_i }$ by Claim~\ref{clm:EOQ_properties}. The next claim, whose proof is provided in Section~\ref{subsec:proof_lem_easy_regime_EOQ}, shows that when $T_{\min}^* \geq \frac{ n }{ \eps } \cdot \frac{ K_0 }{ \widetilde{\opt} }$, this policy happens to be near-optimal. 

\begin{lemma} \label{lem:easy_regime_EOQ}
When $T_{\min}^* \geq \frac{ n }{ \eps } \cdot \frac{ K_0 }{ \widetilde{\opt} }$, we have
\[ F(\hat{T}) ~~\leq~~ \sum_{i \in [n]} C_i^+( \hat{T}_i ) ~~\leq~~ (1 + 2\eps) \cdot F(T^*) \ . \]
\end{lemma}

It is important to point out that, since the value of $T_{\min}^*$ is unknown from an algorithmic perspective, we have no way of telling whether the optimal policy $T^*$ falls within the easy regime or not. Consequently, the above-mentioned policy $\hat{T}$ will be kept aside as a candidate solution that will compete against the policy we construct in subsequent sections for the difficult regime. Eventually, the better out of these two policies will be our final solution. Along these lines, yet another important remark is that we can go around an explicit evaluation of the long-run average cost $F(\hat{T})$. Without additional arguments, this operation requires a direct calculation of the joint ordering cost $J( \hat{T} )$ via the closed-form expression in Lemma~\ref{lem:limit_joint_order}, which potentially involves $\Omega( 2^n )$ distinct non-zero terms. However, Lemma~\ref{lem:easy_regime_EOQ} shows that $F(\hat{T}) \leq \sum_{i \in [n]} C_i^+( \hat{T}_i ) \leq (1 + 2\eps) \cdot F(T^*)$, meaning that we can instead evaluate the straightforward term $\sum_{i \in [n]} C_i^+( \hat{T}_i )$.

\paragraph{The difficult regime: $\bs{T_{\min}^* < \frac{ n }{ \eps } \cdot \frac{ K_0 }{ \widetilde{\opt} }}$.} We proceed by further observing that $T_{\min}^* \geq \frac{ K_0 }{ \widetilde{\opt} }$; this bound can easily be derived by noting that
\[ \frac{ K_0 }{ T_{\min}^* } ~~\leq~~ J(T^*) ~~\leq~~ F(T^*) ~~\leq~~ \widetilde{\opt} \ . \]
As such, in the difficult regime, we know that $T_{\min}^*$ resides within the interval $[\frac{ K_0 }{ \widetilde{\opt} }, \frac{ n }{ \eps } \cdot \frac{ K_0 }{ \widetilde{\opt} })$, whose endpoints differ by a multiplicative factor of $\frac{ n }{ \eps }$. This property allows us to assume that we have at our possession an under-estimate $\tilde{T}_{\min}$ of the minimal ordering interval $T_{\min}^*$, specifically, one that satisfies
\begin{equation} \label{eqn:rel_tildeTmin_Tmin}
\left( 1 - \frac{ \eps }{ 2 } \right) \cdot T_{\min}^* ~~\leq~~ \tilde{T}_{\min} ~~\leq~~ T_{\min}^* \ . 
\end{equation}
The latter assumption can be enforced by testing all powers of $1 + \frac{ \eps }{ 2 }$ within $[\frac{ K_0 }{ \widetilde{\opt} }, \frac{ n }{ \eps } \cdot \frac{ K_0 }{ \widetilde{\opt} })$ as candidate values for $\tilde{T}_{\min}$, noting that there are only $O( \frac{ 1 }{ \eps } \log ( \frac{ n }{ \eps } ) )$ such values. With this estimate in place, Sections~\ref{subsec:alignment_intuition}-\ref{subsec:policy_description} describe how our algorithmic approach operates in the difficult regime, leaving its cost analysis to be discussed in Section~\ref{sec:analysis}.

\subsection{High-level plan} \label{subsec:alignment_intuition}

The fundamental intuition behind $\Psi$-pairwise alignment can be succinctly stated as follows: Our objective is to efficiently construct a small-sized set of representative points $\hat{\cal R} \subseteq \bbR_+$, guaranteed to simultaneously be $\eps$-dense and $\eps$-assignable. To better understand the last few properties, it is instructive to keep in mind the following interpretation:
\begin{enumerate}
\item {\em Small size:} By a small-sized set of representatives, $\hat{\cal R}$, we mean that its cardinality is upper-bounded as a function of $\frac{ 1 }{ \eps }$ and nothing more.

\item {\em $\eps$-density:} The set $\hat{\cal R}$ will be called $\eps$-dense when, by placing joints order at all integer multiples of all representative points, we obtain an ordering density $\lim_{\Delta \to \infty} \frac{ N(\hat{\cal R},\Delta) }{ \Delta }$  that matches the analogous density $\lim_{\Delta \to \infty} \frac{ N(T^*,\Delta) }{ \Delta }$ with respect to the optimal policy $T^*$, up to a factor of $1 + \eps$. Of course, by representation~\eqref{eqn:joint_cost_def}, this property translates to $J( \hat{\cal R} ) \leq (1 + \eps) \cdot J( T^* )$, implying that our long-run joint ordering cost is near-optimal.

\item {\em $\eps$-assignability:} We say that $\hat{\cal R}$ is $\eps$-assignable when, for each commodity $i \in [n]$, we can choose an integer multiple of some representative in $\hat{\cal R}$ to serve as the ordering interval $\hat{T}_i$ of this commodity, such that its marginal operating cost $C_i( \hat{T}_i )$ is within factor $1 + \eps$ of the analogous cost $C_i( T_i^* )$ with respect to $T^*$.
\end{enumerate}
It is worth mentioning that, at the moment, even the mere existence of representatives that concurrently satisfy properties 1-3 is unclear. For instance, $0$-density and $0$-assignability can clearly be ensured by setting  $\hat{\cal R} = \{ T_1^*, \ldots, T_n^* \}$. However, a scenario where we end up with $| \hat{\cal R} | = \Omega( n )$ cannot be ruled out, meaning that regardless of its computational aspects, this construction simply fails to obtain a small-sized set.
  
\subsection{Preliminaries: Segments, activity, and representatives} \label{subsec:alg_definitions}

\paragraph{Interval classification.} As our first step toward an efficient construction of the form described in Section~\ref{subsec:alignment_intuition}, let us say that an ordering interval $T_i^*$ is large when $T_i^* > \frac{ 1 }{ \eps } \cdot \tilde{T}_{\min}$. In the opposite case, inequality~\eqref{eqn:rel_tildeTmin_Tmin} implies that $T_i^* \in [\tilde{T}_{\min}, \frac{ 1 }{ \eps } \cdot \tilde{T}_{\min}]$, in which case this interval will be referred to as being small. We further partition $[\tilde{T}_{\min}, \frac{ 1 }{ \eps } \cdot \tilde{T}_{\min}]$ by powers of $1 + \eps$ into the sequence of segments $S_1, \ldots, S_L$, such that   
\[ S_1 ~~=~~ [\tilde{T}_{\min}, (1 + \eps) \cdot \tilde{T}_{\min}), \quad S_2 ~~=~~ [(1 + \eps) \cdot\tilde{T}_{\min}, (1 + \eps)^2 \cdot \tilde{T}_{\min}), \quad \ldots \]
so on and so forth, where in general $S_{\ell} = [(1 + \eps)^{\ell-1} \cdot\tilde{T}_{\min}, (1 + \eps)^{\ell} \cdot \tilde{T}_{\min})$, noting that $L = O( \frac{ 1 }{ \eps } \log \frac{ 1 }{ \eps } )$. 

\paragraph{Guessing active segments and defining representatives.} We say that the segment $S_{\ell}$ is active with respect to the optimal policy $T^*$ when there is at least one commodity $i \in [n]$ with $T_i^* \in S_{\ell}$. Let ${\cal A}^* \subseteq [L]$ be the index set of active segments, which is clearly unknown from an algorithmic perspective. We begin by guessing the precise identity of ${\cal A}^*$, or equivalently, whether each segment is active or not with respect to $T^*$. For this purpose, the overall number of guesses to consider is $2^L = O( 2^{ O( \frac{ 1 }{ \eps } \log \frac{ 1 }{ \eps } ) } )$. Next, for each active segment $S_{\ell}$, let $R^*_{\ell}$ be an arbitrarily picked interval $T_i^*$ that belong to this segment. We refer to $R^*_{\ell}$ as the representative of $S_{\ell}$, noting that the exact value of 
$R^*_{\ell}$ is clearly unknown and that this definition is meant to serve only for purposes of analysis. 
 
\paragraph{Basic properties.} We proceed by listing two useful observations regarding the set of representatives ${\cal R}^* = \{ R^*_{\ell} \}_{ \ell \in {\cal A}^* }$. First, Observation~\ref{obs:rel_NRstar_NTstar} informs us that, for every $\Delta \geq 0$, the number of joint orders in $[0,\Delta]$ with respect to the time intervals ${\cal R}^*$ is upper-bounded by the analogous quantity with respect to the optimal policy $T^*$; this claim can be straightforwardly inferred by noting that ${\cal R}^* \subseteq \{ T_1^*, \ldots, T_n^* \}$. In addition, Observation~\ref{obs:Rstar1_in_Rstar} implicitly states that $S_1$ must be an active segment, meaning that its representative $R_1^*$ belongs to ${\cal R}^*$. To verify this claim, it suffices to note that $T_{\min}^* \in [\tilde{T}_{\min}, (1 + \eps) \cdot \tilde{T}_{\min}) = S_1$, by inequality~\eqref{eqn:rel_tildeTmin_Tmin}.

\begin{observation} \label{obs:rel_NRstar_NTstar}
$N( {\cal R}^*, \Delta ) \leq N( T^*, \Delta )$, for every $\Delta \geq 0$.
\end{observation}

\begin{observation} \label{obs:Rstar1_in_Rstar}
$R_1^* \in {\cal R}^*$.
\end{observation}

\subsection{\texorpdfstring{$\bs{\Psi}$}{}-pairwise alignment} \label{subsec:alg_alignment} 

Moving forward, our goal would be to efficiently compute a small-sized, $\eps$-dense, and $\eps$-assignable set of representative $\hat{\cal R}$ by approximately  mimicking how the optimal set ${\cal R}^*$ is structured. Specifically, we wish to synchronize the least common multiples of certain subsets of representatives that will eventually be shown to determine the joint ordering cost function $J(\cdot)$ up to lower-order terms. 

\paragraph{Alignment guessing.} For this purpose, we say that a pair of active segments $S_{\ell_1}$ and $S_{\ell_2}$ is aligned when their representatives $R^*_{\ell_1}$ and $R^*_{\ell_2}$ have common integer multiples, which is  equivalent to $\frac{ R^*_{\ell_1} }{ R^*_{\ell_2} }$ being a rational number. Furthermore, letting $M_{\ell_1, \ell_2}^*$ be the least common multiple of $R^*_{\ell_1}$ and $R^*_{\ell_2}$, this pair of segments is called $\Psi$-aligned when the corresponding multiples $\frac{ M_{\ell_1, \ell_2}^* }{ R^*_{\ell_1} }$ and $\frac{ M_{\ell_1, \ell_2}^* }{ R^*_{\ell_2} }$ both take values of at most $\Psi$, with the latter constant set to $\Psi = \frac{ 2 L^2 \cdot 2^L }{ \eps } = O( 2^{ O( \frac{ 1 }{ \eps } \log \frac{ 1 }{ \eps } ) } )$. In this case, we make use of $\alpha_{ \{ \ell_1, \ell_2 \}, \ell_1 }$ and $\alpha_{ \{ \ell_1, \ell_2 \}, \ell_2 }$ to denote these two multiples, respectively. Moreover, ${\cal P}_{\Psi}^*$ will stand for the collection of $\Psi$-aligned pairs. 

Given these definitions, our next algorithmic step consists of guessing the set ${\cal P}_{\Psi}^*$. Put differently, for each pair of active segments $S_{\ell_1}$ and $S_{\ell_2}$, we guess whether they are $\Psi$-aligned or not, amounting to $O( 2^{ O( |{\cal A}^*|^2 ) } ) = O( 2^{ O( \frac{ 1 }{ \eps^2 } \log^2 \frac{ 1 }{ \eps } ) } )$ options overall. For each such pair, we additionally guess $\alpha_{ \{ \ell_1, \ell_2 \}, \ell_1 }$ and $\alpha_{ \{ \ell_1, \ell_2 \}, \ell_2 }$; here, the total number of guesses is $O( \Psi^{ O( |{\cal A}^*|^2 ) } ) = O( 2^{ O( \frac{ 1 }{ \eps^3 } \log^3 \frac{ 1 }{ \eps } ) } )$.

\paragraph{Obtaining approximate representatives.} With these ingredients in place, our method for synchronizing common multiples, at least with respect to certain subsets of representatives, utilizes the following linear feasibility problem: 
\begin{equation} \label{eqn:LP_representatives}
\tag{LP}
\begin{array}{ll}
(A)~~R_{\ell} \in \bar{S}_{\ell} & \forall \, \ell \in {\cal A}^* \\
(B)~~\alpha_{ \{ \ell_1, \ell_2 \}, \ell_1 } \cdot R_{\ell_1} = \alpha_{ \{ \ell_1, \ell_2 \}, \ell_2 } \cdot R_{\ell_2} \qquad & \forall \, (\ell_1,\ell_2) \in {\cal P}_{\Psi}^* 
\end{array}
\end{equation}
Here, for each active segment $S_{\ell}$, we introduce a decision variable $R_{\ell}$, playing the role of its so-called approximate representative. Constraint~(A) states that each representative is picked within its corresponding segment. As a side note, since the right endpoint of $S_{\ell} = [(1 + \eps)^{\ell-1} \cdot\tilde{T}_{\min}, (1 + \eps)^{\ell} \cdot \tilde{T}_{\min})$ induces a strict inequality, to define a valid linear program, we plug in the closed segment $\bar{S}_{\ell} = [(1 + \eps)^{\ell-1} \cdot\tilde{T}_{\min}, (1 + \eps)^{\ell} \cdot \tilde{T}_{\min}]$ instead. Constraint~(B) ensures that every $\Psi$-aligned pair $S_{\ell_1}$ and $S_{\ell_2}$ has its representatives $R_{\ell_1}$ and $R_{\ell_2}$ related through the same multiples that relate  the optimal representatives $R_{\ell_1}^*$ and $R_{\ell_2}^*$. It is worth mentioning that some pairs could be aligned but not $\Psi$-aligned, in which case there are no guarantees on how their common multiples are synchronized. Moreover, constraint~(B) involves pairs and nothing more, implying that \eqref{eqn:LP_representatives} does not include direct guarantees for arbitrary subsets of three or more representatives.

\paragraph{Feasibility of \eqref{eqn:LP_representatives}.} Let us first observe that the linear program \eqref{eqn:LP_representatives} is indeed feasible, since $\{ R_{\ell}^* \}_{ \ell \in {\cal A}^* }$ forms a feasible solution, by construction. In order to solve this problem, one can supposedly employ any black-box polynomial-time procedure for linear optimization, noting that every extreme point solution can be specified via polynomially-many bits. To verify this claim, it suffices to observe that every entry in the constraint matrix of \eqref{eqn:LP_representatives} is integer-valued, bounded by 
$\Psi = O( 2^{ O( \frac{ 1 }{ \eps } \log \frac{ 1 }{ \eps } ) } )$. Similarly, every entry in its right-hand-side vector is either zero, or a rational number residing within $[\tilde{T}_{\min}, \frac{ 1 }{ \eps } \cdot \tilde{T}_{\min}]$. That said, in Section~\ref{subsec:structure_feasible}, we propose a graph-based method to obtain carefully-constructed feasible solutions to \eqref{eqn:LP_representatives}, allowing us to enforce several structural properties that will be imperative to our subsequent analysis.

\subsection{The structure of feasible solutions} \label{subsec:structure_feasible}

\paragraph{The alignment graph $\bs{G_{\Psi}^*}$ and separability by component.} Recalling that ${\cal P}_{\Psi}^*$ designates the collection of $\Psi$-aligned pairs of active segments, let us consider the undirected graph $G_{\Psi}^* = ({\cal A}^*, {\cal P}_{\Psi}^*)$. Here, the vertex set is comprised of the active segments, and each pair of such segments is connected by an edge when they are $\Psi$-aligned. We make use of $\cc( G_{\Psi}^* )$ to denote the underlying collection of connected components in this graph. 

The important observation to take note of is that the linear formulation \eqref{eqn:LP_representatives} is separable by component, since there are no constraints tying between representatives from two different components. For this reason, we can focus our attention on independently solving the single-component programs $\{ \text{\eqref{eqn:LP_per_component}} \}_{C \in \cc( G_{\Psi}^* )}$, where
\begin{equation} \label{eqn:LP_per_component}
\tag{LP$_{C}$}
\begin{array}{ll}
(A)~~R_{\ell} \in \bar{S}_{\ell} & \forall \, \ell \in C \\
(B)~~\alpha_{ \{ \ell_1, \ell_2 \}, \ell_1 } \cdot R_{\ell_1} = \alpha_{ \{ \ell_1, \ell_2 \}, \ell_2 } \cdot R_{\ell_2} \qquad & \forall \, (\ell_1,\ell_2) \in {\cal P}_{\Psi}^* : \ell_1,\ell_2 \in C
\end{array}
\end{equation}

\paragraph{\eqref{eqn:LP_per_component} is a single-variable problem.} Focusing on a single component $C$, let $\sigma_C$ be an arbitrarily picked vertex in $C$, to which we refer as the source of this component. Roughly speaking, the upcoming claim states that in any feasible solution, once the approximate representative $R_{\sigma_C}$ is fixed, the values of all other representatives are uniquely determined in a solution-independent way. The finer details of this claim are formalized in Lemma~\ref{lem:scaling_of_source}, whose proof is provided in Section~\ref{subsec:proof_lem_scaling_of_source}. Interestingly, our proof is constructive, showing how to compute the undermentioned coefficients in $O( ( \frac{ 1 }{ \eps } )^{ O(1) } )$ time.

\begin{lemma} \label{lem:scaling_of_source}
There exists a collection of coefficients $\{ \beta_{\ell} \}_{\ell \in C}$ satisfying the next two properties:
\begin{enumerate}
\item In any feasible solution to \eqref{eqn:LP_per_component}, we have $R_{\ell} = \beta_{\ell} \cdot R_{ \sigma_C}$ for every $\ell \in C$.

\item For every $\ell \in C$, we have $\beta_{\ell} = \frac{ \beta_{\ell}^+ }{ \beta_{\ell}^- }$, where $\beta_{\ell}^+$ and $\beta_{\ell}^-$ are integers bounded by $\Psi^{ |C| }$.
\end{enumerate}
\end{lemma}

\paragraph{Feasibility region of \eqref{eqn:LP_per_component}.} An immediate consequence of this structural result is that, in order to obtain a feasible solution to each of the single-component programs $\{ \text{\eqref{eqn:LP_per_component}} \}_{C \in \cc( G_{\Psi}^* )}$, there is no need to employ general-purpose linear optimization tools. Instead, for each component $C$, we should simply identify the approximate representative $R_{\sigma_C}$ of its source $\sigma_C$ with respect to some feasible solution, exploiting the constructive proof of Lemma~\ref{lem:scaling_of_source} to uniquely determine all other representatives. Along these lines, since constraints~(A) and~(B) prescribe a non-empty polyhedral set, we know that the range of possible values for $R_{\sigma_C}$ forms a closed subsegment, say $[r_{ \sigma_C }^-, r_{ \sigma_C }^+] \subseteq \bar{S}_{\sigma_C}$. It is not difficult to verify that the left endpoint $r_{ \sigma_C }^-$ is precisely the smallest value of $R_{\sigma_C}$ for which $R_{\ell} = \beta_{\ell} \cdot R_{ \sigma_C}$ resides within $\bar{S}_{\ell}$, for every $\ell \in C$. Similarly, the right endpoint $r_{ \sigma_C }^+$ is given by the largest value of $R_{\sigma_C}$ that satisfies this condition. Based on these observations, both $r_{ \sigma_C }^-$ and $r_{ \sigma_C }^+$ can be computed in $O( ( \frac{ 1 }{ \eps } )^{ O(1) } )$ time.

\paragraph{Preventing coincidental alignments.} While any value in $[r_{ \sigma_C }^-, r_{ \sigma_C }^+]$ can play the role of $R_{\sigma_C}$ in terms of being a feasible solution to \eqref{eqn:LP_per_component}, this choice will be made in a very specific way. Intuitively, for every pair of segments $S_{\ell_1}$ and $S_{\ell_2}$ in different connected components, we wish to ensure that their corresponding representatives $R_{\ell_1}$ and $R_{\ell_2}$ are misaligned, in the sense that their joint multiples occur very infrequently in comparison to those of $\min \{ R_{\ell_1}, R_{\ell_2} \}$. 
We formalize the latter notion by asking that $\alpha_1 \cdot R_{\ell_1} \neq \alpha_2 \cdot R_{\ell_2}$ for every pair of integers $\alpha_1 \leq \Psi$ and $\alpha_2 \leq \Psi$, noting that this property may be violated by arbitrary  solutions to~\eqref{eqn:LP_representatives}. That said, in the next claim, whose proof is given in Section~\ref{subsec:proof_lem_feas_sol_misalign}, we show that misalignment is indeed achievable.

\begin{lemma} \label{lem:feas_sol_misalign}
We can construct a feasible solution to \eqref{eqn:LP_representatives} in which every pair of representatives belonging to different components is misaligned. Our construction can be implemented in $O( 2^{ O( \frac{ 1 }{ \eps } \log \frac{ 1 }{ \eps } ) } )$ time.
\end{lemma}

\paragraph{Approximate representatives vs.\ optimal ones.} Yet another important consequence of the preceding discussion is more analytical in nature, being crucial for a number of cost accounting arguments in Section~\ref{sec:analysis}. As previously mentioned, the optimal representatives $\{ R_{\ell}^* \}_{ \ell \in {\cal A}^* }$ form a feasible solution to 
\eqref{eqn:LP_representatives}, meaning in turn that their restriction to each connected component $C$ is feasible with respect to \eqref{eqn:LP_per_component}. As such, for every feasible solution $\{ R_{\ell} \}_{ \ell \in {\cal A}^* }$ and for every component $C$, we know that both $R_{\sigma_C}^*$ and $R_{\sigma_C}$ reside within $[r_{ \sigma_C }^-, r_{ \sigma_C }^+]$, implying that there is a solution-dependent coefficient $\gamma_{C,R} \in [ \frac{ r_{ \sigma_C }^- }{ r_{ \sigma_C }^+ }, \frac{ r_{ \sigma_C }^+ }{ r_{ \sigma_C }^- } ]$ for which $R_{\sigma_C} = \gamma_{C,R} \cdot R^*_{\sigma_C}$. However, Lemma~\ref{lem:scaling_of_source}(1) extends this property to the entire component, as formally stated in Corollary~\ref{cor:scaling_component} below. Here, we further observe that, since $[ \frac{ r_{ \sigma_C }^- }{ r_{ \sigma_C }^+ }, \frac{ r_{ \sigma_C }^+ }{ r_{ \sigma_C }^- } ] \subseteq \bar{S}_{\sigma_C}$ and since the endpoints of the latter segment differ by a multiplicative factor of $1 + \eps$, we know that $\gamma_{C,R} \in 1 \pm \eps$.

\begin{corollary} \label{cor:scaling_component}
For every feasible solution $\{ R_{\ell} \}_{ \ell \in {\cal A}^* }$ to 
\eqref{eqn:LP_representatives} and for every component $C \in \cc( G_{\Psi}^* )$, there exists a coefficient $\gamma_{C,R} \in 1 \pm \eps$ such that $R_{\ell} = \gamma_{C,R} \cdot R^*_{\ell}$ for every $\ell \in C$.
\end{corollary}

\subsection{The final policy} \label{subsec:policy_description}

We are now ready to lay down the specifics of our replenishment policy, which will be denoted by $\hat{T} = (\hat{T}_1, \ldots, \hat{T}_n)$. To this end, given a feasible solution $\hat{\cal R} = \{ \hat{R}_{\ell} \}_{ \ell \in {\cal A}^* }$ to the linear formulation \eqref{eqn:LP_representatives}, constructed according to Lemma~\ref{lem:feas_sol_misalign}, we proceed in two steps:
\begin{enumerate}
\item {\em Placing joint orders:} Joint orders will be placed only at integer multiples of the approximate representatives $\{ \hat{R}_{\ell} \}_{ \ell \in {\cal A}^* }$. In other words, with ${\cal M}_{\hat{R}_{\ell}} = \{ 0, \hat{R}_{\ell}, 2\hat{R}_{\ell}, \ldots \}$ standing for the integer multiples of $\hat{R}_{\ell}$, we decide in advance to open a joint order at every point in $\bigcup_{ \ell \in {\cal A}^* } {\cal M}_{\hat{R}_{\ell}}$, regardless of whether any given point will subsequently be utilized by some commodity or not.

\item {\em Placing commodity-specific orders:} For each commodity $i \in [n]$, we determine its ordering interval $\hat{T}_i$ to be the one that minimizes its marginal EOQ-based cost $C_i(\cdot)$ out of the following options:
\begin{itemize}
\item {\em Small intervals:} Any of the approximate representatives $\{ \hat{R}_{\ell} \}_{ \ell \in {\cal A}^* }$.

\item {\em Single large interval:} Letting $T^{\max}_i = \max \{ \frac{ 1 }{ \eps } \cdot \tilde{T}_{\min}, \sqrt{ K_i / H_i  } \}$, the additional option is $\lceil T^{\max}_i \rceil^{ (\hat{R}_1) }$, where $\lceil \cdot \rceil^{ (\hat{R}_1) }$ is an operator that rounds its argument up to the nearest integer multiple of $\hat{R}_1$.
\end{itemize}
\end{enumerate}
It is important to emphasize that, while choosing one of the above-mentioned ``small'' intervals as the ordering interval $\hat{T}_i$ clearly falls within our set of joint orders, this also happens to be the case for the ``large'' interval option. Indeed, by Observation~\ref{obs:Rstar1_in_Rstar}, we know that $\hat{R}_1 \in \hat{\cal R}$, implying that ordering commodity $i$ according to the interval $\lceil T^{\max}_i \rceil^{ (\hat{R}_1) }$ falls on integer multiples of $\hat{R}_1$, where joint orders have already been placed.
\section{Analysis} \label{sec:analysis}

The upcoming contents will be dedicated to pinpointing the performance guarantee of our resulting policy, showing that its long-run average cost is within factor $1 + \eps$ of optimal. To this end, following the high-level outline of Section~\ref{subsec:alignment_intuition}, our analysis proceeds as follows:
\begin{itemize}
    \item {\em Establishing $O(\eps)$-density:} Recalling that the replenishment policy $\hat{T}$ places joint orders at integer multiples of the approximate representatives $\{ \hat{R}_{\ell} \}_{ \ell \in {\cal A}^* }$, we argue in Section~\ref{subsec:cost_joint_orders} that the latter set is $5\eps$-dense. Namely, its ordering density $\lim_{\Delta \to \infty} \frac{ N(\hat{\cal R},\Delta) }{ \Delta }$ will be upper-bounded by $(1 + 5\eps) \cdot \lim_{\Delta \to \infty} \frac{ N(T^*,\Delta) }{ \Delta }$, where $T^*$ stands for the optimal replenishment policy. By representation~\eqref{eqn:joint_cost_def}, this property directly implies that our long-run joint ordering cost is near-optimal, in the sense that $J( \hat{T} ) \leq (1 + 5\eps) \cdot J( T^* )$.

    \item {\em Establishing $O(\eps)$-assignability:} Concurrently, we prove in Section~\ref{subsec:cost_comm_orders} that the set of representatives $\{ \hat{R}_{\ell} \}_{ \ell \in {\cal A}^* }$ is in fact $2\eps$-assignable, meaning that our choice of the ordering intervals $\hat{T}_1, \ldots, \hat{T}_n$ 
    guarantees $C_i( \hat{T}_i ) \leq (1 + 2 \eps) \cdot C_i( T_i^* )$, for every commodity $i \in [n]$. In other words, we show that the marginal operating cost of each commodity with respect to the approximate policy $\hat{T}$ is within factor $1 + 2\eps$ of the analogous quantity with respect to the optimal policy $T^*$.

    \item {\em Evaluating long-run average operating costs:} Finally, since our algorithmic approach employs numerous guessing steps, to ultimately identify the least expensive policy out of all possible outcomes, it is imperative to efficiently evaluate the long-run cost function $F( \cdot )$ for each of the resulting policies. As explained prior to deriving Lemma~\ref{lem:limit_joint_order}, we do not know how to accomplish this goal for arbitrarily-structured policies. However, Section~\ref{subsec:evaluation_F} is devoted to showing that the unique structural properties of our particular policies can be leveraged to estimate their long-run cost within any degree of accuracy.
\end{itemize}

\subsection{Cost analysis: Joint orders} \label{subsec:cost_joint_orders}

\paragraph{Intent.} Let $\hat{\cal R} = \{ \hat{R}_{\ell} \}_{ \ell \in {\cal A}^* }$ be the collection of approximate representatives, constructed along the lines of Section~\ref{sec:algorithm}. In what follows, we relate the ordering density of this set to that of the optimal policy $T^*$, specifically arguing that
\begin{equation} \label{eqn:lim_NhatT_NTstar}
\lim_{\Delta \to \infty} \frac{ N(\hat{\cal R},\Delta) }{ \Delta } ~~\leq~~ (1 + 5\eps) \cdot \lim_{\Delta \to \infty} \frac{ N(T^*,\Delta) }{ \Delta } \ . 
\end{equation}
To this end, it is worth mentioning that $N( {\cal R}^*, \Delta ) \leq N( T^*, \Delta )$ for every $\Delta \geq 0$, by Observation~\ref{obs:rel_NRstar_NTstar}. Therefore, inequality~\eqref{eqn:lim_NhatT_NTstar} can be inferred from the next claim, whose proof is presented in the remainder of this section.

\begin{lemma} \label{lem:UB_points_hatR}
For every $\Delta \geq 0$, we have
\[ N(\hat{\cal R},\Delta) ~~\leq~~  (1 + 5\eps) \cdot N({\cal R}^*,\Delta) + 5 \cdot | {\cal A}^* |^2 \cdot 2^{ |{\cal A}^* | } \ . \]    
\end{lemma}

\paragraph{Inclusion-exclusion with (un)crossing sets.} We remind the reader that $N(\hat{\cal R},\Delta)$ stands for the number of joint orders in $[0,\Delta]$ with respect to the ordering intervals $\hat{\cal R}$.  Letting ${\cal M}_{\hat{R}_{\ell},\Delta} = \{ 0, \hat{R}_{\ell}, 2\hat{R}_{\ell}, \ldots, \lfloor \frac{ \Delta }{ \hat{R}_{\ell} } \rfloor \cdot \hat{R}_{\ell} \}$ be the set of integer multiples of $\hat{R}_{\ell}$ within $[0,\Delta]$, we clearly have
\[ N(\hat{\cal R},\Delta) ~~=~~ \left| \bigcup_{\ell \in {\cal A}^*} {\cal M}_{\hat{R}_{\ell},\Delta} \right| \ . \] 
Let us classify a subset of segments ${\cal A} \subseteq {\cal A}^*$ as uncrossing when it is contained in a single connected component of the alignment graph $G_{\Psi}^*$. In the opposite case, where ${\cal A}$ has segments belonging to two or more components, we say that ${\cal A}$ is crossing. These two families of sets will be respectively designated by $\bar{\cal X}$ and ${\cal X}$. With this terminology, by the inclusion-exclusion principle, $N(\hat{\cal R},\Delta)$ can be written as:
\begin{eqnarray}
N(\hat{\cal R},\Delta) & = & \sum_{{\cal A} \subseteq {\cal A}^*} \left| \bigcap_{\ell \in {\cal A}} {\cal M}_{\hat{R}_{\ell},\Delta} \right| \cdot (-1)^{ |{\cal A}| + 1} \nonumber \\
& = & \underbrace{ \sum_{ {\cal A} \in \bar{\cal X} } \left| \bigcap_{\ell \in {\cal A}} {\cal M}_{\hat{R}_{\ell},\Delta} \right| \cdot (-1)^{ |{\cal A}| + 1} }_{ \text{(I) uncrossing} } + \underbrace{ \sum_{ {\cal A} \in {\cal X} } \left| \bigcap_{\ell \in {\cal A}} {\cal M}_{\hat{R}_{\ell},\Delta} \right| \cdot (-1)^{ |{\cal A}| + 1} }_{ \text{(II) crossing} } \ . \label{eqn:UB_NhatR_I_II}
\end{eqnarray}
We proceed by separately upper-bounding the terms (I) and (II) appearing in this expression.

\paragraph{Bounding the crossing term (II).} Starting with the easier part, we argue in the next claim that the crossing term (II) is actually negligible in comparison to the number of joints orders $N({\cal R}^*,\Delta)$ with respect to the optimal representatives ${\cal R}^*$. At least intuitively, this claim stems from the observation that crossing sets necessarily contain at least one misaligned pair of representatives, making their joint orders very infrequent in comparison to the ordering density of an optimal replenishment policy. We formalize this notion while establishing the next upper bound.

\begin{lemma} \label{lem:UB_term_II}
$\mathrm{(II)} \leq  \eps \cdot N({\cal R}^*,\Delta) + 2^{ |{\cal A}^*| }$.
\end{lemma}
\begin{proof}
Let us first focus on a single crossing set ${\cal A} \in {\cal X}$, with $| \bigcap_{\ell \in {\cal A}} {\cal M}_{\hat{R}_{\ell},\Delta} |$ being its corresponding term within the overall summation (II). Since ${\cal A}$ is crossing, this set has a pair of segments $S_{\ell_1}$ and $S_{\ell_2}$ in different connected components of $G_{\Psi}^*$. As a result, by Lemma~\ref{lem:feas_sol_misalign}, we know that $\hat{R}_{\ell_1}$ and $\hat{R}_{\ell_2}$ either do not have common integer multiples, or have their least common multiple being greater than $\Psi \cdot \min \{ \hat{R}_{\ell_1}, \hat{R}_{\ell_2} \}$. Therefore,
\begin{eqnarray}
\left| \bigcap_{\ell \in {\cal A}} {\cal M}_{\hat{R}_{\ell},\Delta} \right| & \leq &  \left| {\cal M}_{\hat{R}_{\ell_1},\Delta} \cap {\cal M}_{\hat{R}_{\ell_2},\Delta} \right| \nonumber \\
& \leq & \left\lfloor \frac{ \Delta }{ \Psi \cdot \min \{ \hat{R}_{\ell_1}, \hat{R}_{\ell_2} \} } \right\rfloor + 1 \nonumber \\
& \leq & \frac{ \Delta }{  \Psi \cdot \tilde{T}_{\min} } + 1 \nonumber \\
& \leq & \frac{ 2 }{ \Psi } \cdot N({\cal R}^*,\Delta) + 1 \ . \label{eqn:UB_intersect_cross}
\end{eqnarray}
Here, the third inequality holds since both $\hat{R}_{\ell_1}$ and $\hat{R}_{\ell_2}$ reside within the interval $[\tilde{T}_{\min}, \frac{ 1 }{ \eps } \cdot \tilde{T}_{\min}]$, due to being part of a feasible solution to \eqref{eqn:LP_representatives}. The fourth inequality is obtained by noting that according to inequality~\eqref{eqn:rel_tildeTmin_Tmin}, we have in particular $( 1 - \frac{ \eps }{ 2 } ) \cdot T_{\min}^* \leq \tilde{T}_{\min}$, and thus $N({\cal R}^*,\Delta) \geq \frac{ \Delta }{ T_{\min}^* } \geq \frac{ \Delta }{  2\tilde{T}_{\min} }$. It is important to emphasize that, for precisely the same argument,  inequality~\eqref{eqn:UB_intersect_cross} is valid with respect to ${\cal R}^*$, meaning that $| \bigcap_{\ell \in {\cal A}} {\cal M}_{R^*_{\ell},\Delta} | \leq \frac{ 2 }{ \Psi } \cdot N({\cal R}^*,\Delta) + 1$; we will make use of this property later on.

Given these observations, the crossing term (II) can be bounded by noting that 
\begin{eqnarray}
\mathrm{(II)} & = & \sum_{ {\cal A} \in {\cal X} } \left| \bigcap_{\ell \in {\cal A}} {\cal M}_{\hat{R}_{\ell},\Delta} \right| \cdot (-1)^{ |{\cal A}| + 1} \nonumber \\
& \leq & \sum_{ {\cal A} \in {\cal X} } \left| \bigcap_{\ell \in {\cal A}} {\cal M}_{\hat{R}_{\ell},\Delta} \right| \label{eqn:bound_II_midway} \\
& \leq & 2^{ |{\cal A}^*| } \cdot \left(  \frac{ 2 }{ \Psi } \cdot N({\cal R}^*,\Delta) + 1 \right)\nonumber \\
& \leq & \eps \cdot N({\cal R}^*,\Delta) + 2^{ |{\cal A}^*| } \ , \nonumber
\end{eqnarray}
where the second inequality follows from~\eqref{eqn:UB_intersect_cross}, and the third inequality holds since $\Psi = \frac{ 2 L^2 \cdot 2^L }{ \eps } \geq \frac{ 2 \cdot |{\cal A}^* |^2 \cdot 2^{ |{\cal A}^* | } }{ \eps }$, as explained in Section~\ref{subsec:alg_alignment}.
\end{proof}

\paragraph{Bounding the uncrossing term (I).} We now shift our attention to the more difficult part, claiming that the uncrossing term (I) does not deviate much above the number of joints orders $N({\cal R}^*,\Delta)$. This analysis is precisely where most structural properties instilled by $\Psi$-pairwise alignment will play an instrumental role, and it is therefore advisable to keep in mind the main bullet points of Section~\ref{subsec:structure_feasible}.

\begin{lemma} \label{lem:UB_term_I}
$\mathrm{(I)} \leq (1 + 4\eps) \cdot N({\cal R}^*,\Delta) + 4 \cdot | {\cal A}^* |^2 \cdot 2^{ |{\cal A}^* | }$.
\end{lemma}
\begin{proof}
Similarly to how the proof of Lemma~\ref{lem:UB_term_II} starts off, let us initially consider a single uncrossing set ${\cal A} \in \bar{\cal X}$, with $| \bigcap_{\ell \in {\cal A}} {\cal M}_{\hat{R}_{\ell},\Delta} |$ being its corresponding term within the overall summation (I). Our intermediate objective is to relate this term to the analogous quantity $| \bigcap_{\ell \in {\cal A}} {\cal M}_{R^*_{\ell},\Delta} |$, defined with respect to ${\cal R}^*$. For this purpose, letting $C \in \cc( G_{\Psi}^* )$ be the connected component of $G_{\Psi}^*$ that contains ${\cal A}$, we know by Corollary~\ref{cor:scaling_component} that there exists a coefficient $\gamma_{C,\hat{\cal R}} \in 1 \pm \eps$ such that $\hat{R}_{\ell} = \gamma_{C,\hat{\cal R}} \cdot R^*_{\ell}$ for every $\ell \in C$. In the next claim, we exploit this connection to show that the latter coefficient determines the relation between $| \bigcap_{\ell \in {\cal A}} {\cal M}_{\hat{R}_{\ell},\Delta} |$ and 
$| \bigcap_{\ell \in {\cal A}} {\cal M}_{R^*_{\ell},\Delta} |$ up to constant factors. The proof of this result appears in Section~\ref{subsec:proof_clm_bound_good_set}. 

\begin{claim} \label{clm:bound_good_set}
$| \bigcap_{\ell \in {\cal A}} {\cal M}_{\hat{R}_{\ell},\Delta} | \in \frac{ 1 }{ \gamma_{C, \hat{R}} } \cdot | \bigcap_{\ell \in {\cal A}} {\cal M}_{R^*_{\ell},\Delta} | \pm 2$.
\end{claim}

Consequently, the uncrossing term (I) can be bounded by observing that 
\begin{eqnarray*}
\mathrm{(I)} & = & \sum_{ {\cal A} \in \bar{\cal X} } \left| \bigcap_{\ell \in {\cal A}} {\cal M}_{\hat{R}_{\ell},\Delta} \right| \cdot (-1)^{ |{\cal A}| + 1}  \\
& = & \sum_{C \in \cc( G_{\Psi}^* )} \sum_{{\cal A} \subseteq C}  \left| \bigcap_{\ell \in {\cal A}} {\cal M}_{\hat{R}_{\ell},\Delta} \right| \cdot (-1)^{ |{\cal A}| + 1}  \\
& \leq & \sum_{C \in \cc( G_{\Psi}^* )} \frac{ 1 }{ \gamma_{C,\hat{\cal R}} } \cdot  \sum_{{\cal A} \subseteq C} \left| \bigcap_{\ell \in {\cal A}} {\cal M}_{R^*_{\ell},\Delta} \right|  \cdot (-1)^{ |{\cal A}| + 1} + 2 \cdot 2^{ |{\cal A}^*| }  \\
& = & \sum_{C \in \cc( G_{\Psi}^* )} \frac{ 1 }{ \gamma_{C,\hat{\cal R}} } \cdot   \left| \bigcup_{\ell \in C} {\cal M}_{R^*_{\ell},\Delta} \right| + 2 \cdot 2^{ |{\cal A}^*| }  \\
& \leq & (1 + 2\eps) \cdot \underbrace{ \sum_{C \in \cc( G_{\Psi}^* )} \left| \bigcup_{\ell \in C} {\cal M}_{R^*_{\ell},\Delta} \right| }_{ \text{(III)} } + 2 \cdot 2^{ |{\cal A}^*| } \ , 
\end{eqnarray*}
where the first inequality is implied by Claim~\ref{clm:bound_good_set}, and the second holds since $\gamma_{C,\hat{\cal R}} \geq 1 - \eps$ and $\eps \in (0,\frac{1}{2})$. The next claim, whose proof is provided in Section~\ref{subsec:proof_clm_bound_sum_comp}, shows that the resulting summation $\mathrm{(III)}$ cannot exceed $N({\cal R}^*,\Delta)$ by much, thereby concluding the overall proof.

\begin{claim} \label{clm:bound_sum_comp}
$\mathrm{(III)} \leq (1 + \eps) \cdot N({\cal R}^*,\Delta) + | {\cal A}^* |^2 \cdot 2^{ |{\cal A}^* | }$.
\end{claim}    
\end{proof}

\paragraph{Putting it all together.} We are now ready to finalize the proof of Lemma~\ref{lem:UB_points_hatR}. To this end, by plugging Lemmas~\ref{lem:UB_term_II} and~\ref{lem:UB_term_I} into decomposition~\eqref{eqn:UB_NhatR_I_II}, it indeed follows that
\[ N(\hat{\cal R},\Delta) ~~=~~ \mathrm{(I)} + \mathrm{(II)} ~~\leq~~ (1 + 5\eps) \cdot N({\cal R}^*,\Delta) + 5 \cdot | {\cal A}^* |^2 \cdot 2^{ |{\cal A}^* | } \ . \]

\subsection{Cost analysis: Commodity-specific orders} \label{subsec:cost_comm_orders}

\paragraph{Intent.} Let $\hat{T} = (\hat{T}_1, \ldots, \hat{T}_n)$ be the replenishment policy constructed in Section~\ref{sec:algorithm}. We proceed by arguing that, for every commodity $i \in [n]$, its marginal operating cost with respect to the approximate policy $\hat{T}$ matches the analogous quantity with respect to the optimal policy $T^*$ up to low-order terms. The precise nature of this relation is formalized in Lemma~\ref{lem:UB_marginal_hatT}, whose proof is presented in the remainder of this section. Our analysis will be divided to three parametric regimes, depending on whether the optimal interval $T_i^*$ is small or large and on how $T_i^*$ and $\sqrt{ K_i / H_i }$ are related. 

\begin{lemma} \label{lem:UB_marginal_hatT}
$C_i( \hat{T}_i ) \leq (1 + 2 \eps) \cdot C_i( T_i^* )$, for every commodity $i \in [n]$.   
\end{lemma}

\paragraph{Regime 1: $\bs{T_i^*}$ is small.} Let $\ell \in [L]$ be the unique segment index for which $T_i^* \in S_{\ell}$, meaning in particular that this segment is active. Since $\hat{\cal R}$ is a feasible solution to \eqref{eqn:LP_representatives}, it follows that the decision variable $\hat{R}_{\ell}$ indeed exists within this linear program, and moreover, $\hat{R}_{\ell} \in \bar{S}_{\ell}$ due to constraint~(A). Therefore, $\hat{R}_{\ell}$ is one of the options considered for our ordering interval $\hat{T}_i$, as explained in Section~\ref{subsec:policy_description}. Moreover, since we pick the option that minimizes the marginal EOQ-based cost $C_i(\cdot)$ of this commodity, 
\begin{eqnarray*}
C_i( \hat{T}_i ) & \leq & C_i( \hat{R}_{\ell} ) \\
& = & \frac{ K_i }{ \hat{R}_{\ell} } + H_i \hat{R}_{\ell} \\
& \leq & (1 + \eps) \cdot \left( \frac{ K_i }{ T_i^* } + H_i T_i^* \right) \\
& = & (1 + \eps) \cdot C_i( T_i^* ) \ .
\end{eqnarray*}
Here, the second inequality holds since both $T_i^*$ and $\hat{R}_{\ell}$ reside within $\bar{S}_{\ell}$, which is a segment whose endpoints differ by a factor of $1+\eps$, implying that $\frac{ T^*_i }{ 1+\eps } \leq \hat{R}_{\ell} \leq (1 + \eps) \cdot T_i^*$.

\paragraph{Regime 2: $\bs{T_i^*}$ is large and $\bs{T^*_i \leq \sqrt{ K_i / H_i }}$.} By recalling how large ordering intervals were defined in Section~\ref{subsec:alg_definitions}, we must have $\frac{ 1 }{ \eps } \cdot \tilde{T}_{\min} < T_i^* \leq \sqrt{ K_i / H_i }$ in this case, implying that $T^{\max}_i = \max \{ \frac{ 1 }{ \eps } \cdot \tilde{T}_{\min}, \sqrt{ K_i / H_i } \} = \sqrt{ K_i / H_i }$. Since $\lceil T^{\max}_i \rceil^{ (\hat{R}_1) }$ is one of the options considered in setting our ordering interval $\hat{T}_i$, and since we pick the option that minimizes the marginal cost function $C_i(\cdot)$, it follows that
\begin{eqnarray}
C_i( \hat{T}_i ) & \leq & C_i( \lceil T^{\max}_i \rceil^{ (\hat{R}_1) } ) \nonumber \\
& = & C_i( \lceil \sqrt{ K_i / H_i } \rceil^{ (\hat{R}_1) } ) \nonumber \\
& \leq & C_i( (1 + 2\eps) \cdot \sqrt{ K_i / H_i } ) \label{eqn:comm_analysis_case2_1} \\
& = & \frac{ 1 }{ 2 } \cdot \left( 1 + 2\eps + \frac{ 1 }{ 1 + 2\eps } \right) \cdot C_i( \sqrt{ K_i / H_i } ) \label{eqn:comm_analysis_case2_2} \\
& \leq & (1 + \eps) \cdot C_i( T_i^* ) \ . \nonumber
\end{eqnarray}
Here, inequality~\eqref{eqn:comm_analysis_case2_1} holds since the function $C_i(\cdot)$ is strictly convex, with a unique minimum at $\sqrt{ K_i / H_i }$, as stated in items~1 and~2 of Claim~\ref{clm:EOQ_properties}. Therefore, this function is strictly increasing over $[\sqrt{ K_i / H_i }, \infty)$, and the desired inequality follows by noting that 
\begin{eqnarray*}
\left\lceil \sqrt{ K_i / H_i } \right\rceil^{ (\hat{R}_1) } & \leq & \sqrt{ K_i / H_i } + \hat{R}_1 \\
& \leq & \sqrt{ K_i / H_i } + (1 + \eps) \cdot \tilde{T}_{\min} \\
& \leq & (1 + 2\eps) \cdot \sqrt{ K_i / H_i } \ ,
\end{eqnarray*}
where the second and third inequalities are respectively obtained by noting that $\hat{R}_1 \in \bar{S}_1 = [\tilde{T}_{\min}, (1 + \eps) \cdot \tilde{T}_{\min}]$ and $\tilde{T}_{\min} < \eps \cdot \sqrt{ K_i / H_i }$, as explained above. Finally, equality~\eqref{eqn:comm_analysis_case2_2} is precisely the well-known scaling property of the optimal EOQ solution, formally stated in item~3 of Claim~\ref{clm:EOQ_properties}.

\paragraph{Regime 3: $\bs{T_i^*}$ is large and $\bs{T^*_i > \sqrt{ K_i / H_i } }$.} Once again, since $\lceil T^{\max}_i \rceil^{ (\hat{R}_1) }$ is one of the options considered for our ordering interval $\hat{T}_i$, we infer that
\begin{eqnarray*}
C_i( \hat{T}_i ) & \leq & C_i( \lceil T^{\max}_i \rceil^{ (\hat{R}_1) } ) \\
& \leq & C_i( (1 + 2\eps) \cdot T_i^* ) \\
& = & \frac{ K_i }{ (1 + 2\eps) \cdot T_i^* } + (1 + 2\eps) \cdot H_i T_i^* \\
& \leq & (1 + 2\eps) \cdot C_i( T_i^* ) \ .
\end{eqnarray*}
Here, the second inequality holds since, as explained when analyzing regime~2, the function $C_i(\cdot)$ is strictly increasing over $[\sqrt{ K_i / H_i }, \infty)$. In addition,
$T^{\max}_i \geq \sqrt{ K_i / H_i }$, and the desired inequality follows by noting that
\begin{eqnarray*}
\lceil T^{\max}_i \rceil^{ (\hat{R}_1) } & \leq & T^{\max}_i + \hat{R}_1 \\
& \leq & T^{\max}_i + (1 + \eps) \cdot \tilde{T}_{\min} \\
& \leq & (1 + 2\eps) \cdot T^{\max}_i \\
& \leq & (1 + 2\eps) \cdot T_i^* \ , 
\end{eqnarray*}
where the third and fourth inequalities are obtained by recalling that $T^*_i \geq \max \{ \frac{ 1 }{ \eps } \cdot \tilde{T}_{\min}, \sqrt{ K_i / H_i } \} = T^{\max}_i$, according to the case hypothesis of regime~3. 

\subsection{Evaluating long-run average costs} \label{subsec:evaluation_F}

A close examination of our algorithmic approach reveals that, due to a sequence of enumeration-based procedures required for guessing the  minimal ordering interval $T_{\min}^*$, the set of active segments ${\cal A}^*$, and the $\alpha_{\{\cdot,\cdot\},\cdot}$-multiples, we have generated a collection of $ O( 2^{ O( \frac{ 1 }{ \eps^3 } \log^3 \frac{ 1 }{ \eps } ) } \cdot \log n )$ candidate policies overall. Thus, it remains to efficiently evaluate the long-run cost $F( \hat{T} ) =  J(\hat{T}) + \sum_{i \in [n]} C_i( \hat{T}_i )$ for each of these policies and identify the least expensive outcome.

While calculating $\sum_{i \in [n]} C_i( \hat{T}_i )$ is straightforward, it appears as if we would run into an exponential-time computation when evaluating the joint ordering cost $J(\hat{T})$. Indeed, by Lemma~\ref{lem:limit_joint_order}, the latter function can be expressed as $J(\hat{T}) = K_0 \cdot \sum_{{\cal N} \subseteq [n]} \frac{ (-1)^{ |{\cal N}| + 1 } }{ M_{\hat{T}, {\cal N}} }$, where $M_{\hat{T}, {\cal N}}$ is the least common multiple of $\{ \hat{T}_i \}_{i \in {\cal N}}$, with the convention that $M_{\hat{T}, {\cal N}} = \infty$ when these intervals do not have common multiples. However, as explained in Section~\ref{subsec:policy_description}, our policy $\hat{T}$ places joint orders only at integer multiples of the approximate representatives $\hat{\cal R} = \{ \hat{R}_{\ell} \}_{ \ell \in {\cal A}^* }$. This feature allows us to establish the next estimate for the joint ordering cost, showing that it is essentially determined by common multiples of uncrossing sets. The proof of this result is deferred to Section~\ref{subsec:proof_lem_short_exp_joint_hatR}. 

\begin{lemma} \label{lem:short_exp_joint_hatR}
$J( \hat{\cal R} ) \in (1 \pm 2\eps) \cdot K_0 \cdot \sum_{ {\cal A} \in \bar{\cal X} } \frac{ (-1)^{ |{\cal A}| + 1} }{ M_{\hat{R}, {\cal A}} }$.
\end{lemma}

Evidently, the approximate expression we have just obtained for evaluating $J( \hat{\cal R} )$ involves only $O( 2^{ |{\cal A}^* | } ) =O( 2^{ O(\frac{ 1 }{ \eps } \log \frac{ 1 }{ \eps }) } )$ summands. Moreover, since each of the sets ${\cal A} \subseteq C$ is uncrossing, its approximate representatives $\{ \hat{R}_{\ell} \}_{ \ell \in {\cal A} }$ have common integer multiples, as explained within the proof of Claim~\ref{clm:bound_good_set}, with the additional observation that $(\prod_{\ell \in {\cal A}} \beta_{\ell}^+ ) \cdot \hat{R}_{ \sigma_C}$ divides by $\hat{R}_{\ell} =  \frac{ \beta_{\ell}^+ }{ \beta_{\ell}^- } \cdot \hat{R}_{ \sigma_C}$, for every $\ell \in {\cal A}$. In turn, the least common multiple $M_{\hat{R}, {\cal A}}$ must be of the form $\frac{ \beta^+_{\cal A} }{ \beta^-_{\cal A} } \cdot \hat{R}_{ \sigma_C}$, where both $\beta^+_{\cal A}$ and $\beta^-_{\cal A}$ are integers bounded by $\Psi^{ |C| } = O( 2^{ O( \frac{ 1 }{ \eps^2 } \log^2 \frac{ 1 }{ \eps } ) } )$. As such, $M_{\hat{R}, {\cal A}}$ can easily be computed by enumerating over all possible values of $\beta^+_{\cal A}$ and $\beta^-_{\cal A}$.
\section{Additional Proofs} \label{sec:more_proofs}

\subsection{Proof of Lemma~\ref{lem:limit_joint_order}} \label{subsec:proof_lem_limit_joint_order}

To establish the desired claim, recalling that ${\cal M}_{T_i,\Delta} = \{ 0, T_i, 2T_i, \ldots, \lfloor \frac{ \Delta }{ T_i } \rfloor \cdot T_i \}$, by the inclusion-exclusion principle we have
\begin{eqnarray*}
N(T,\Delta) & = & \left| \bigcup_{i \in [n]} {\cal M}_{T_i,\Delta} \right| \\
& = & \sum_{{\cal N} \subseteq [n]} \left| \bigcap_{i \in {\cal N}}{\cal M}_{T_i,\Delta} \right| \cdot (-1)^{ |{\cal N}| + 1} \\
& = & \sum_{{\cal N} \subseteq [n]} \left( \left\lfloor \frac{ \Delta }{ M_{\cal N} } \right\rfloor + 1 \right) \cdot (-1)^{ |{\cal N}| + 1} \ .
\end{eqnarray*}
Consequently, $N(T,\Delta) \in \Delta \cdot \sum_{{\cal N} \subseteq [n]} \frac{ (-1)^{ |{\cal N}| + 1 } }{ M_{\cal N} } \pm 2^n$, and by the squeeze theorem it follows that
\[ \lim_{\Delta \to \infty} \frac{ N(T,\Delta) }{ \Delta } ~~=~~ \sum_{{\cal N} \subseteq [n]} \frac{ (-1)^{ |{\cal N}| + 1 } }{ M_{\cal N} } \ . \]

\subsection{Proof of Lemma~\ref{lem:easy_regime_EOQ}} \label{subsec:proof_lem_easy_regime_EOQ}

We begin by showing that the first inequality, $F(\hat{T}) \leq \sum_{i \in [n]} C_i^+( \hat{T}_i )$ is always valid, regardless of the parametric regime in question and regardless of the policy being considered. To this end, note that
\begin{eqnarray*}
F( \hat{T} ) & = & J( \hat{T}) + \sum_{i \in [n]} C_i( \hat{T}_i )  \\
& \leq & \sum_{i \in [n]} \left( \frac{ K_0 + K_i }{ \hat{T}_i } + H_i \hat{T}_i \right)  \\
& = & \sum_{i \in [n]} C_i^+( \hat{T}_i  )  \ .
\end{eqnarray*}
To better understand the inequality above, letting ${\cal M}_{\hat{T}_i,\Delta} = \{ 0, \hat{T}_i, 2\hat{T}_i, \ldots, \lfloor \frac{ \Delta }{ \hat{T}_i } \rfloor \cdot T_i \}$ be the set of integer multiples of $\hat{T}_i$ within $[0,\Delta]$, by representation~\eqref{eqn:joint_cost_def} of the joint ordering cost, we have
\[ \frac{ J( \hat{T} ) }{ K_0 } ~~=~~ \lim_{\Delta \to \infty} \frac{ N(\hat{T},\Delta) }{ \Delta } ~~=~~ \lim_{\Delta \to \infty} \frac{ | \bigcup_{i \in [n]} {\cal M}_{\hat{T}_i,\Delta} | }{ \Delta }  ~~\leq~~ \lim_{\Delta \to \infty}   \sum_{i \in [n]} \frac{ | {\cal M}_{\hat{T}_i,\Delta} | }{ \Delta } ~~=~~ \sum_{i \in [n]} \frac{ 1 }{ \hat{T}_i } \ . \]

We proceed to establish the second inequality, $\sum_{i \in [n]} C_i^+( \hat{T}_i ) \leq (1 + 2\eps) \cdot F(T^*)$, whose validity very much depends on being in the easy regime and on the specific choice of the policy $\hat{T}$. For this purpose, recalling that for each commodity $i \in [n]$, the ordering interval $\hat{T}_i$ minimizes the overloaded cost function $C_i^+( \cdot )$, we have in particular
\begin{eqnarray*}
\sum_{i \in [n]} C_i^+( \hat{T}_i ) & \leq & \sum_{i \in [n]} C_i^+( T_i^* )  \\
&= & \sum_{i \in [n]} \left( \frac{ K_0 + K_i }{ T^*_i } + H_i T^*_i \right)  \\
& \leq & \frac{ n K_0 }{ T_{\min}^* } + \sum_{i \in [n]} C_i( T^*_i )  \\
& \leq & \eps \cdot \widetilde{\opt} + F( T^*) \\
& \leq & (1 + 2\eps) \cdot F(T^*) \ . 
\end{eqnarray*}
Here, the last two inequalities respectively hold since $T_{\min}^* \geq \frac{ n }{ \eps } \cdot \frac{ K_0 }{ \widetilde{\opt} }$ and $\widetilde{\opt} \leq 2 \cdot F(T^*)$.

\subsection{Proof of Lemma~\ref{lem:scaling_of_source}} \label{subsec:proof_lem_scaling_of_source}

To establish the desired claim via an efficient construction, let us pick an arbitrary spanning tree ${\cal T}$ within the connected component $C$. Focusing on a single vertex $\ell \in C$, we make use of ${\cal T}_{\sigma_C,\ell}$ to denote the unique path in ${\cal T}$ connecting the source $\sigma_C$ to this vertex. Now, suppose that $\sigma_C = u_1, \ldots, u_k = \ell$ is the sequence of vertices along this path. We first observe that since $(u_1, u_2)$ is an edge of the alignment graph $G_{\Psi}^*$, constraint~(B) of \eqref{eqn:LP_per_component} forces us to set $\alpha_{ \{ u_1, u_2 \}, u_1 } \cdot R_{u_1} = \alpha_{ \{ u_1, u_2 \}, u_2 } \cdot R_{u_2}$ for this particular pair. Similarly, since
$(u_2, u_3)$ is an edge of $G_{\Psi}^*$, this constraint sets $\alpha_{ \{ u_2, u_3 \}, u_2 } \cdot R_{u_2} = \alpha_{ \{ u_2, u_3 \}, u_3 } \cdot R_{u_3}$. Letting this observation propagate throughout the entire path ${\cal T}_{\sigma_C,\ell}$, its resulting sequence of equations can be aggregated to obtain a unique value for the representative $R_{\ell}$, given by:
\[ R_{\ell} ~~=~~ R_{u_k} ~~=~~ \left( \prod_{\kappa \in [k-1]} \frac{ \alpha_{ \{ u_{\kappa}, u_{\kappa+1} \}, u_{\kappa} } }{ \alpha_{ \{ u_{\kappa}, u_{\kappa+1} \}, u_{\kappa+1} } } \right) \cdot R_{u_1} ~~=~~ \left( \prod_{\kappa \in [k-1]} \frac{ \alpha_{ \{ u_{\kappa}, u_{\kappa+1} \}, u_{\kappa} } }{ \alpha_{ \{ u_{\kappa}, u_{\kappa+1} \}, u_{\kappa+1} } } \right) \cdot R_{\sigma_C} \ . \]

Therefore, in any feasible solution to \eqref{eqn:LP_per_component}, once the approximate representative $R_{\sigma_C}$ is fixed, we have just shown that $R_{\ell} = \beta_{\ell} \cdot R_{ \sigma_C}$ for every $\ell \in C$, where $\beta_{\ell}$ corresponds to the above-mentioned solution-independent coefficient, $\prod_{\kappa \in [k-1]} \frac{ \alpha_{ \{ u_{\kappa}, u_{\kappa+1} \}, u_{\kappa} } }{ \alpha_{ \{ u_{\kappa}, u_{\kappa+1} \}, u_{\kappa+1} } }$. It is easy to verify that the collection of coefficients $\{ \beta_{\ell} \}_{\ell \in C}$ can be computed in $O( |C|^{O(1)} ) = O( ( \frac{ 1 }{ \eps } )^{ O(1) } )$ time. Moreover, since each of the  multiples $\alpha_{ \{ \cdot, \cdot \}, \cdot }$ takes a value of at most $\Psi$, as explained in Section~\ref{subsec:alg_alignment}, it follows that $\beta_{\ell}$ can be expressed as a ratio of the form $\frac{ \beta_{\ell}^+ }{ \beta_{\ell}^- }$, where $\beta_{\ell}^+$ and $\beta_{\ell}^-$ are integers bounded by $\Psi^{ |C| }$.

\subsection{Proof of Lemma~\ref{lem:feas_sol_misalign}} \label{subsec:proof_lem_feas_sol_misalign}

To end up with a feasible solution, it suffices to ensure that, for every connected component $C \in \cc( G_{\Psi}^* )$, the approximate representative $R_{\sigma_C}$ of its source $\sigma_C$ resides within $[r_{ \sigma_C }^-, r_{ \sigma_C }^+]$. Toward making a specific choice in this interval, we say that component $C$ is tight when $r_{ \sigma_C }^- = r_{ \sigma_C }^+$; otherwise, this component will be referred to as being loose.

Starting with the tight case, for any such component $C$, there is  only one possible value for the representative of its source, $R_{\sigma_C} = r_{ \sigma_C }^- = r_{ \sigma_C }^+$. In turn, this choice uniquely determines the value of every other representative in $C$, by Lemma~\ref{lem:scaling_of_source}. Given these decisions, we argue that for every pair of segments $S_{\ell_1}$ and $S_{\ell_2}$ in different tight  components, say $C_1$ and $C_2$, their representatives $R_{\ell_1}$ and $R_{\ell_2}$ must be misaligned. To verify this claim,
recalling that the optimal representatives $\{ R_{\ell}^* \}_{ \ell \in {\cal A}^* }$ form a feasible solution to 
\eqref{eqn:LP_representatives}, we necessarily have $R_{\sigma_{C_1}} = R_{\sigma_{C_1}}^*$ and $R_{\sigma_{C_2}} = R_{\sigma_{C_2}}^*$, implying that $R_{\ell_1} = R_{\ell_1}^*$ and $R_{\ell_2} = R_{\ell_2}^*$, again by Lemma~\ref{lem:scaling_of_source}. However, since $S_{\ell_1}$ and $S_{\ell_2}$ belong to different   components, we know that they are not $\Psi$-aligned, meaning that $\alpha_1 \cdot R_{\ell_1}^* \neq \alpha_2 \cdot R_{\ell_2}^*$ for every pair of integers $\alpha_1 \leq \Psi$ and $\alpha_2 \leq \Psi$, which is of course equivalent to $\alpha_1 \cdot R_{\ell_1} \neq \alpha_2 \cdot R_{\ell_2}$.

Moving on to consider the loose case, we iterate through these components one after the other, in arbitrarily order. When each such component $C$ is examined, the important observation is that, for every already-fixed component $\hat{C}$, either tight or loose, there are only $O( |C| \cdot |\hat{C}| \cdot \Psi^2 )$ values in $[r_{ \sigma_C }^-, r_{ \sigma_C }^+]$ that would create an alignment between a pair of representatives, one in $C$ and the other in $\hat{C}$. Indeed, each such value corresponds to solving $\alpha_1 \cdot R_{\ell_1} = \alpha_2 \cdot R_{\ell_2}$ for some pair of segments $\ell_1 \in C$ and $\ell_2 \in \hat{C}$, and for some pair of integers $\alpha_1 \leq \Psi$ and $\alpha_2 \leq \Psi$. Consequently, there are $O( |{\cal A}^*|^2 \cdot  \Psi^2 )$ easily-recognizable values to avoid in this interval, and we can guarantee that alignments will not be created by picking any other option, noting that $r_{ \sigma_C }^- < r_{ \sigma_C }^+$.

\subsection{Proof of Claim~\ref{clm:bound_good_set}} \label{subsec:proof_clm_bound_good_set}

We begin by noting that, since ${\cal A}$ is an uncrossing set, the approximate representatives $\{ \hat{R}_{\ell} \}_{ \ell \in {\cal A} }$ must have common integer multiples. To ascertain this claim, by circling back to Lemma~\ref{lem:scaling_of_source},
one can easily 
verify that $(\prod_{\ell \in {\cal A}} \beta_{\ell}^+ ) \cdot \hat{R}_{ \sigma_C}$ divides by $\hat{R}_{\ell} =  \frac{ \beta_{\ell}^+ }{ \beta_{\ell}^- } \cdot \hat{R}_{ \sigma_C}$, for every $\ell \in {\cal A}$. Moreover, since these arguments apply to any feasible solution, the optimal representatives $\{ R_{\ell}^* \}_{ \ell \in {\cal A} }$ also have common integer multiples. Given this observation, we make use of $\hat{M}_{\cal A}$ and $M^*_{\cal A}$ to respectively denote the least common multiples of $\{ \hat{R}_{\ell} \}_{ \ell \in {\cal A} }$ and $\{ R^*_{\ell} \}_{ \ell \in {\cal A} }$. In order to relate between these quantities, we remind the reader that $\hat{R}_{\ell} = \gamma_{C,\hat{\cal R}} \cdot R^*_{\ell}$ for every $\ell \in C$, implying in turn that $\hat{M}_{\cal A} = \gamma_{C,\hat{\cal R}} \cdot M^*_{\cal A}$. We can now derive the desired claim by noting that
\begin{eqnarray*}
\left| \bigcap_{\ell \in {\cal A}} {\cal M}_{\hat{R}_{\ell},\Delta} \right| & = & \left\lfloor \frac{ \Delta }{ \hat{M}_{\cal A} } \right\rfloor + 1 \\
& \leq & \frac{ \Delta }{ \gamma_{C,\hat{\cal R}} \cdot M^*_{\cal A} }  + 1 \\
& \leq & \frac{ 1 }{ \gamma_{C,\hat{\cal R}} } \cdot \left( \left\lfloor \frac{ \Delta }{  M^*_{\cal A} } \right\rfloor + 1 \right)  + 1 \\
& = & \frac{ 1 }{ \gamma_{C,\hat{\cal R}} } \cdot \left| \bigcap_{\ell \in {\cal A}} {\cal M}_{R^*_{\ell},\Delta} \right| + 1 \ . 
\end{eqnarray*}
A nearly identical sequence of inequalities in the opposite direction shows that $| \bigcap_{\ell \in {\cal A}} {\cal M}_{\hat{R}_{\ell},\Delta} | \geq \frac{ 1 }{ \gamma_{C,\hat{\cal R}} } \cdot | \bigcap_{\ell \in {\cal A}} {\cal M}_{R^*_{\ell},\Delta} | - 2$.

\subsection{Proof of Claim~\ref{clm:bound_sum_comp}} \label{subsec:proof_clm_bound_sum_comp}

We begin by introducing a decomposition of $N({\cal R}^*,\Delta)$ into the contributions of different connected components, keeping in mind that the latter term designates the number of joint orders in $[0,\Delta]$ with respect to the ordering intervals ${\cal R}^*$. To this end, letting $U_C^* = \bigcup_{\ell \in C} {\cal M}_{R^*_{\ell},\Delta}$ be the set of such orders with respect to the representatives of each component $C \in \cc( G_{\Psi}^* )$, we can rewrite 
$N({\cal R}^*,\Delta)$ by observing that
\begin{eqnarray*}
 N({\cal R}^*,\Delta) & = & \left| \bigcup_{\ell \in {\cal A}^*} {\cal M}_{R^*_{\ell},\Delta} \right| \\
 & = & \left| \bigcup_{C \in \cc( G_{\Psi}^*)} U_C^* \right| \\
& = & \sum_{B \subseteq \cc( G_{\Psi}^*)} \left| \bigcap_{C \in B} U_C^* \right| \cdot (-1)^{ |B|+1} \\
& = & \sum_{ C \in \cc( G_{\Psi}^*) } \left| U_C^* \right| + \sum_{ \MyAbove{ B \subseteq \cc( G_{\Psi}^*): }{ |B| \geq 2 }} \left| \bigcap_{C \in B} U_C^* \right| \cdot (-1)^{ |B|+1} \\
& = & \mathrm{(III)} + \underbrace{ \sum_{ \MyAbove{ B \subseteq \cc( G_{\Psi}^*): }{ |B| \geq 2 }} \left| \bigcap_{C \in B} U_C^* \right| \cdot (-1)^{ |B|+1} }_{ \text{(IV)} } \ .
\end{eqnarray*}
Therefore, to establish the desired bound on (III), it suffices to show that $\mathrm{(IV)} \geq - ( \eps \cdot N({\cal R}^*,\Delta) + | {\cal A}^* |^2 \cdot 2^{ |{\cal A}^* | } )$.

For this purpose, consider some subset $B \subseteq \cc( G_{\Psi}^*)$ with $|B| \geq 2$. Letting $C_1 \neq C_2$ be two components in $B$, we have
\begin{eqnarray*}
\left| \bigcap_{C \in B} U_C^* \right| & \leq & \left| U_{C_1}^* \cap U_{C_2}^* \right| \\
& = & \left| \left( \bigcup_{\ell \in C_1} {\cal M}_{R^*_{\ell},\Delta} \right) \cap \left( \bigcup_{\ell \in C_2} {\cal M}_{R^*_{\ell},\Delta} \right) \right| \\
& \leq & \sum_{ \ell_1 \in C_1 } \sum_{ \ell_2 \in C_2 } \left| {\cal M}_{R^*_{\ell_1},\Delta} \cap {\cal M}_{R^*_{\ell_2},\Delta} \right| \\
& \leq &  | {\cal A}^* |^2 \cdot \left( \frac{ 2 }{ \Psi } \cdot N({\cal R}^*,\Delta) + 1 \right) \\
& \leq & \frac{ \eps }{ 2^{ |{\cal A}^* | } } \cdot N({\cal R}^*,\Delta) + | {\cal A}^* |^2 \ .
\end{eqnarray*}
To understand where the third inequality is coming from, the important observation is that, for any pair of segments $\ell_1$ and $\ell_2$ that reside in different connected components of $G_{\Psi}^*$, their corresponding set $\{ \ell_1, \ell_2 \}$ must be of the crossing type. As such, since during the proof of Lemma~\ref{lem:UB_term_II} we noticed that inequality~\eqref{eqn:UB_intersect_cross} is also valid with respect to ${\cal R}^*$, it follows that $| {\cal M}_{R^*_{\ell_1},\Delta} \cap {\cal M}_{R^*_{\ell_2},\Delta} | \leq \frac{ 2 }{ \Psi } \cdot N({\cal R}^*,\Delta) + 1$. The fourth inequality holds since $\Psi = \frac{ 2 L^2 \cdot 2^L }{ \eps } \geq  \frac{ 2 \cdot |{\cal A}^*|^2 \cdot 2^{|{\cal A}^*|} }{ \eps }$. Consequently, we obtain the desired lower bound on (IV) by noting that
\begin{eqnarray*}
\mathrm{(IV)} & = & \sum_{ \MyAbove{ B \subseteq \cc( G_{\Psi}^*): }{ |B| \geq 2 }} \left| \bigcap_{C \in B} U_C^* \right| \cdot (-1)^{ |B|+1} \\
& \geq & - 2^{ | \cc( G_{\Psi}^*) | } \cdot \left( \frac{ \eps }{ 2^{ |{\cal A}^* | } } \cdot N({\cal R}^*,\Delta) + | {\cal A}^* |^2 \right) \\
& \geq & - ( \eps \cdot N({\cal R}^*,\Delta) + | {\cal A}^* |^2 \cdot 2^{ |{\cal A}^* | } ) \ .
\end{eqnarray*}

\subsection{Proof of Lemma~\ref{lem:short_exp_joint_hatR}} \label{subsec:proof_lem_short_exp_joint_hatR}

In order to derive an approximate estimate for $J( \hat{\cal R} ) = K_0 \cdot \lim_{\Delta \to \infty} \frac{ N(\hat{\cal R},\Delta) }{ \Delta }$, we will exploit decomposition~\eqref{eqn:UB_NhatR_I_II}, which partitions the number of joints orders $N(\hat{\cal R},\Delta)$ into the so-called uncrossing and crossing terms, (I) and (II). Specifically, our first step would be that of bounding the crossing term (II) with respect to $N(\hat{\cal R},\Delta)$ in both directions. To this end, it is easy to verify that minor alterations to the proof of Lemma~\ref{lem:UB_term_II} allow us to argue that
\begin{equation} \label{eqn:bound_II_hatR_terms_upper}
\mathrm{(II)} ~~\leq~~ \eps \cdot N(\hat{\cal R},\Delta) + 2^{ |{\cal A}^*| } \ .
\end{equation}
The required modification is nothing more than replacing the last transition in inequality~\eqref{eqn:UB_intersect_cross} by $\frac{ \Delta }{  \Psi \cdot \tilde{T}_{\min} } + 1 \nonumber \leq \frac{ 2 }{ \Psi } \cdot N(\hat{\cal R},\Delta) + 1$; all other arguments remain unchanged. Now, in the opposite direction, it is not difficult to see that an extra minor alteration can be exercised to show that
\begin{equation} \label{eqn:bound_II_hatR_terms_lower}
\mathrm{(II)} ~~\geq~~ -( \eps \cdot N(\hat{\cal R},\Delta) + 2^{ |{\cal A}^*| }) \ .
\end{equation}
Here, one should simply plug in the opposite form of inequality~\eqref{eqn:bound_II_midway}, stating that
\[ \sum_{ {\cal A} \in {\cal X} } \left| \bigcap_{\ell \in {\cal A}} {\cal M}_{\hat{R}_{\ell},\Delta} \right| \cdot (-1)^{ |{\cal A}| + 1} ~~\geq~~ - \sum_{ {\cal A} \in {\cal X} } \left| \bigcap_{\ell \in {\cal A}} {\cal M}_{\hat{R}_{\ell},\Delta} \right| \ . \]

Consequently, by combining inequalities~\eqref{eqn:bound_II_hatR_terms_upper} and~\eqref{eqn:bound_II_hatR_terms_lower}, it follows that $| \mathrm{(II)} | \leq \eps \cdot N(\hat{\cal R},\Delta) + 2^{ |{\cal A}^*| }$. However, since $N(\hat{\cal R},\Delta) = \mathrm{(I)} + \mathrm{(II)}$ by decomposition~\eqref{eqn:UB_NhatR_I_II}, we conclude the proof by observing that 
\begin{eqnarray*}
\lim_{\Delta \to \infty} \frac{ N(\hat{R},\Delta) }{ \Delta } & \in & (1 \pm 2\eps) \cdot \lim_{\Delta \to \infty} \frac{ \mathrm{(I)} }{ \Delta } \\
& = & (1 \pm 2\eps) \cdot  \lim_{\Delta \to \infty} \frac{ 1 }{ \Delta } \cdot \sum_{ {\cal A} \in \bar{\cal X} } \left| \bigcap_{\ell \in {\cal A}} {\cal M}_{\hat{R}_{\ell},\Delta} \right| \cdot (-1)^{ |{\cal A}| + 1} \\
& = & (1 \pm 2\eps) \cdot  \lim_{\Delta \to \infty} \frac{ 1 }{ \Delta } \cdot \sum_{ {\cal A} \in \bar{\cal X} } \frac{ \Delta }{ M_{\hat{R}, {\cal A}} } \cdot (-1)^{ |{\cal A}| + 1} \\
& = & (1 \pm 2\eps) \cdot \sum_{ {\cal A} \in \bar{\cal X} } \frac{ (-1)^{ |{\cal A}| + 1} }{ M_{\hat{R}, {\cal A}} } \ .
\end{eqnarray*}
\section{Concluding Remarks} \label{sec:conclusions}

We conclude this paper with a number of fundamental questions for future research, ranging from seemingly doable to highly non-trivial. These prospective directions take aim at devising more efficient implementations of the core ideas presented in Section~\ref{sec:algorithm}, as well as at examining whether our approximation scheme can be migrated to nearby inventory management models.

\paragraph{Improved implementations?} As stated in Theorem~\ref{thm:main}, the current form of our algorithmic approach leads to an overall running time of $O( 2^{ \tilde{O}(1/\eps^3) } \cdot n^{ O(1) } )$. Even though an exponential dependency on the accuracy level $\eps$ is inevitable, an interesting direction for future work is that of arriving at lower-order exponential terms. One promising idea along these lines begins by observing that the $2^{ \tilde{O}(1/\eps^3) }$-bottleneck resides only within the guessing procedure for the $\alpha_{\{\cdot,\cdot\},\cdot}$-multiples in Section~\ref{subsec:alg_alignment}. Here, each such multiple can be as large as $\Psi$ and we may be required to enumerate over $\Omega( | {\cal A}^* |^2 )$ pairs of segments, ending up with $O( \Psi^{ O( |{\cal A}^*|^2 ) } ) = O( 2^{ O( \frac{ 1 }{ \eps^3 } \log^3 \frac{ 1 }{ \eps } ) } )$ guesses overall. However, a close inspection of the proof of Lemma~\ref{lem:scaling_of_source} reveals that, since we are picking an arbitrary spanning tree within each connected component, every non-tree edge is completely overlooked. For this reason, our guessing procedure can be alternatively implemented by:
\begin{itemize}
    \item Enumerating all possible forests over the  set of vertices ${\cal A}^*$, where by Cayley's formula \citep[pg.~235-240]{AignerZ18}, there are only $| {\cal A}^* |^{ O( | {\cal A}^* | )}$ forests to consider.
    
    \item Guessing the $\alpha_{\{\cdot,\cdot\},\cdot}$-multiples for edges of this forest, of which there are only $O( | {\cal A}^* | )$.
\end{itemize}
Consequently, the total number of guesses becomes $O( ( \Psi \cdot |{\cal A}^*| )^{ O( |{\cal A}^*| ) } ) = O( 2^{ O( \frac{ 1 }{ \eps^2 } \log^2 \frac{ 1 }{ \eps } ) } )$, immediately leading to an $O( 2^{ \tilde{O}(1/\eps^2) } \cdot n^{ O(1) } )$-time implementation.

\paragraph{Extensibility to additional models?} Within the scope of joint replenishment, this paper has been successful at developing $\Psi$-pairwise alignment, a new mechanism for synchronizing multiple EOQ models, and in turn, for efficiently approximating optimal policies within any degree of accuracy. While $\Psi$-pairwise alignment turned out to outperform power-of-$2$ policies in this context, we still do not know whether our methodology can be leveraged to address nearby inventory management models of similar nature. This research direction will be left as an intriguing open question for future work. We believe that the first candidate in line is very likely to be the one-warehouse multi-retailer problem (see, e.g., \cite{Roundy85, MuckstadtR87, LuP94, LeviRSS08, GayonMRS17}), primarily given its structural similarity to the classical joint replenishment model. Additional candidates could be multi-product lot-sizing problems revolving around assembly and distribution systems, where power-of-$2$ polices have been instrumental. Avid readers could delve into the finer details of these settings by consulting the excellent survey of \cite{MuckstadtR93} as well as by going through the elegant analysis of \cite{TeoB01} for several such problems.

\addcontentsline{toc}{section}{Bibliography}
\bibliographystyle{plainnat}
\bibliography{BIB-JRP}


\end{document}